\documentclass[twocolumn,nofootinbib,superscriptaddress,notitlepage,amsmath,amssymb,fleqn,aps,prd,preprintnumbers,10pt]{revtex4-2}
\usepackage{graphicx,listings,xspace,hyperref}
\usepackage[all]{hypcap}
\usepackage{booktabs}
\usepackage{hyperref}

\newcommand{\eps}{\varepsilon}

\hypersetup{hidelinks,
  pdfauthor={Reet Barik,John M. Campbell,J. Taylor Childers,Stefan H{\"o}che,Max Knobbe},
  pdftitle={A portable implementation of one-loop Feynman integrals},
  pdfkeywords={QCD, Perturbation Theory}
}

\newcommand{\affANL}{Argonne National Laboratory, Lemont, IL, 60439, USA}
\newcommand{\affFermilab}{Fermi National Accelerator Laboratory, Batavia, IL 60510, USA}

\begin{document}
\preprint{FERMILAB-PUB-26-0649-T, MCNET-26-25}
\title{A portable implementation of scalar one-loop Feynman integrals}
\author{Reet Barik}\affiliation{\affANL}
\author{John~M.~Campbell}\affiliation{\affFermilab}
\author{J.~Taylor~Childers}\affiliation{\affANL}
\author{Stefan~H{\"o}che}\affiliation{\affFermilab}
\author{Max~Knobbe}\affiliation{\affFermilab}

\begin{abstract}
    We present the first performance-portable implementation of a
    loop integral library for high-energy particle physics. The new
    code is based on the QCDLoop package, which provides an essential
    component in the software stack for perturbative precision
    calculations at collider experiments. Our implementation extends
    QCDLoop to modern computing architectures by means of the Kokkos
    portability framework, which allows to target most existing CPUs
    and GPUs.
\end{abstract}

\maketitle

\section{Introduction}
\label{sec:intro}

Feynman integrals are one of the most important theoretical tools to
compute interaction and decay probabilities in high-energy particle physics
with the help of perturbative quantum field theory~\cite{Smirnov:2004ym,Weinzierl:2022eaz}.
Over the past decades, new algorithms for the reduction of one-loop Feynman
diagrams to a limited set of scalar master integrals have spurred the development
of automated and semi-automated tools, capable of automating the laborious
task of one-loop calculations in the Standard Model or new physics
theories; for an overview see Refs.~\cite{Heinrich:2020ybq,Campbell:2022qmc,Gross:2022hyw}.
They have become an integral part of the suite of numerical simulation programs
needed to harness the full physics potential of the Large Hadron Collider (LHC)
and potential future particle physics experiments~\cite{Huss:2022ful,Huss:2025nlt}.
At their core, these simulations rely on libraries of one-loop scalar integrals that have been developed and refined over time~\cite{vanOldenborgh:1990yc,Hahn:2006qw,Ellis:2007qk,vanHameren:2010cp,Carrazza:2016gav,Denner:2016kdg}.

These developments coincided with efforts to modernize both experimental computing
frameworks and theoretical simulations which typically run on the Worldwide LHC
Computing Grid (WLCG)~\cite{Bird:2005js}. In the early stages of the Large Hadron
Collider, the machines deployed on the WLCG were based solely on CPUs. Persistent
trends towards artificial intelligence and machine learning, as well as efforts
to provide large scale computing resources in a more centralized fashion, have
induced a shift towards GPU computing. It is expected that this will lead to a
scarcity of conventional CPU computing resources in the long term, a fact that
needs to be considered in the development of simulations for future experiments.

We address part of this problem by providing the first GPU implementation of a
library for one-loop scalar integrals. It constitutes a largely platform-independent
port of the well-established package QCDLoop~\cite{Ellis:2007qk,Carrazza:2016gav}.
QCDLoop is built on a complete set of scalar master integrals that have been
thoroughly tested and validated and are used in well established theory tools
such as MCFM~\cite{Campbell:2015qma,Campbell:2019dru}. Making those integrals
accessible to developers of GPU-based matrix element
generators~\cite{Bothmann:2023gew,Hagebock:2025jyk} will shorten the time
to viable computing solutions for the high-luminosity LHC era and beyond.

Modern performance-portable simulation programs are often developed with the
help of portability frameworks. Here, we employ Kokkos~\cite{kokkos,kokkosv1},
which provides support for most current CPU and GPU models. The structure of
our implementation follows that of the C++ version of
QCDLoop~\cite{Carrazza:2016gav}. Due to the large number of compute cores
available on a typical GPU, and the large overhead typically associated
with memory access, we do not implement a caching algorithm for the
results of previous calculations. Input and output of the calculations
is provided using structures of arrays, which allows for easy thread
parallelization. Both double precision and quadruple precision versions
of the routines are provided.

In Sec.~\ref{sec:Background} of this manuscript we give a brief
introduction to the problem of scalar one-loop integrals.
Section~\ref{sec:impl} discusses the implementation of the algorithms
and presents consistency checks and benchmarks of numerical precision.
Section~\ref{sec:perf} discusses the computing performance on various
architectures, and Sec.~\ref{sec:outlook} presents a brief outlook.

\section{Scalar one-loop integrals}
\label{sec:Background}
\begin{figure*}[t]
  \includegraphics[width=\textwidth]{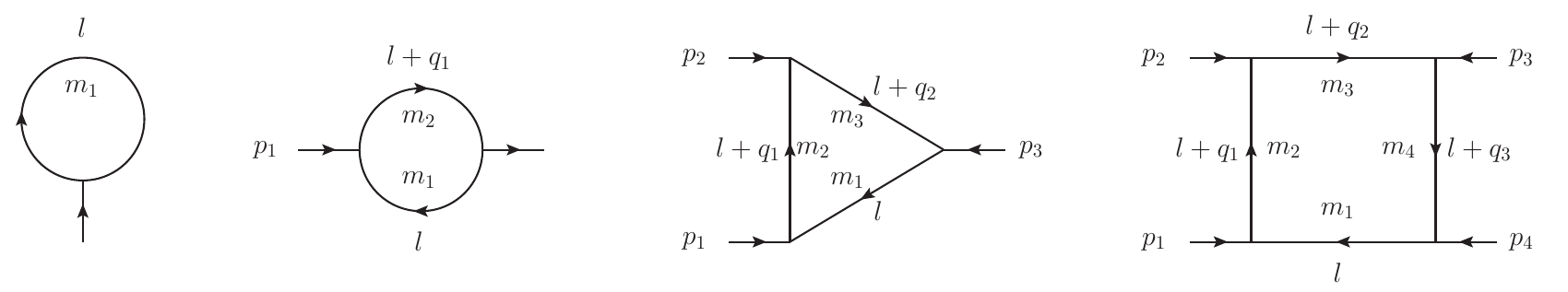}
  \caption{Sketch of the loop integral topologies in
  Eq.~\eqref{eq:topologies}.
  \label{fig:loops}}
\end{figure*}
We use the notation and conventions of Ref.~\cite{Ellis:2007qk}.
The tadpole, bubble, triangle and box integrals are defined as
\begin{widetext}
\begin{equation}\label{eq:topologies}
  \begin{split}
    &I_1^D(m_1^2)=\frac{\mu^{4-D}}{i\pi^{D/2}r_\Gamma}
    \int \frac{d^Dl}{l^2-m_1^2+i\epsilon}\;,\\
    &I_2^D(p_1^2;m_1^2,m_2^2)=\frac{\mu^{4-D}}{i\pi^{D/2}r_\Gamma}
    \int \frac{d^Dl}{(l^2-m_1^2+i\epsilon)((l+q_1)^2-m_2^2+i\epsilon)}\;,\\
    &I_3^D(p_1^2,p_2^2,p_3^2;m_1^2,m_2^2,m_3^2)=\frac{\mu^{4-D}}{i\pi^{D/2}r_\Gamma}
    \int \frac{d^Dl}{(l^2-m_1^2+i\epsilon)((l+q_1)^2-m_2^2+i\epsilon)((l+q_2)^2-m_3^2+i\epsilon)}\;,
  \end{split}
\end{equation}
\begin{equation}\label{eq:topologies_2}
  \begin{split}
    &I_4^D(p_1^2,p_2^2,p_3^2,p_4^2;s_{12},s_{13};m_1^2,m_2^2,m_3^2,m_4^2)=\frac{\mu^{4-D}}{i\pi^{D/2}r_\Gamma}
    \int \frac{d^Dl}{\prod_{n=1}^4((l+q_{n-1})^2-m_n^2+i\epsilon)}\;,
  \end{split}
\end{equation}
\end{widetext}
where $q_n=\sum_{i=1}^np_i$, $q_0=0$, $s_{ij}=(p_i+p_j)^2$, and $D=4-2\eps$.
The topologies described by these integrals are sketched in Fig.~\ref{fig:loops}.
We have defined the normalization factor
\begin{equation}
    r_\Gamma=\frac{\Gamma^2(1-\eps)\Gamma(1+\eps)}{\Gamma(1-2\eps)}\;.
\end{equation}
The results for the loop integrals are typically given in terms of logarithm
and dilogarithm functions and are tabulated in Ref.~\cite{Ellis:2007qk}.
It is important to classify the triangle and box integrals according to their kinematic structure, with the vanishing of internal ($m_i^2$) and external ($p_i^2$) masses controlling the pattern of divergences and leading to different basis integrals. The six basis divergent triangle
integrals, $I_3^D$, are given by
\begin{enumerate}
\item[-] Triangle 1: $I_3^D(0,0,p_3^2;0,0,0)$
\item[-] Triangle 2: $I_3^D(0,p_2^2,p_3^2;0,0,0)$
\item[-] Triangle 3: $I_3^D(0,p_2^2,p_3^2;0,0,m^2)$
\item[-] Triangle 4: $I_3^D(0,p_2^2,m^2;0,0,m^2)$
\item[-] Triangle 5: $I_3^D(0,m^2,m^2;0,0,m^2)$
\item[-] Triangle 6: $I_3^D(m_2^2,s,m_3^2;0,m_2^2,m_3^2)$
\end{enumerate}
The sixteen basis divergent box
integrals, $I_4^D$, are given by
\begin{enumerate}
\item[-] Box 1: $I_4^D(0,0,0,0;s_{12},s_{13};0,0,0,0)$
\item[-] Box 2: $I_4^D(0,0,0,p_4^2;s_{12},s_{13};0,0,0,0)$
\item[-] Box 3: $I_4^D(0,p_2^2,0,p_4^2;s_{12},s_{13};0,0,0,0)$
\item[-] Box 4: $I_4^D(0,0,p_3^2,p_4^2;s_{12},s_{13};0,0,0,0)$
\item[-] Box 5: $I_4^D(0,p_2^2,p_3^2,p_4^2;s_{12},s_{13};0,0,0,0)$
\item[-] Box 6: $I_4^D(0,0,m^2,m^2;s_{12},s_{13};0,0,0,m^2)$
\item[-] Box 7: $I_4^D(0,0,m^2,p_4^2;s_{12},s_{13};0,0,0,m^2)$
\item[-] Box 8: $I_4^D(0,0,p_3^2,p_4^2;s_{12},s_{13};0,0,0,m^2)$
\item[-] Box 9: $I_4^D(0,p_2^2,p_3^2,m^2;s_{12},s_{13};0,0,0,m^2)$
\item[-] Box 10: $I_4^D(0,p_2^2,p_3^2,p_4^2;s_{12},s_{13};0,0,0,m^2)$
\item[-] Box 11: $I_4^D(0,m_3^2,p_3^2,m_4^2;s_{12},s_{13};0,0,m_3^2,m_4^2)$\vspace*{2mm}
\item[-] Box 12: $I_4^D(0,m_3^2,p_3^2,p_4^2;s_{12},s_{13};0,0,m_3^2,m_4^2)$
\item[-] Box 13: $I_4^D(0,p_2^2,p_3^2,p_4^2;s_{12},s_{13};0,0,m_3^2,m_4^2)$
\item[-] Box 14: $I_4^D(m_2^2,m_2^2,m_4^2,m_4^2;s_{12},s_{13};0,m_2^2,0,m_4^2)$
\item[-] Box 15: $I_4^D(m_2^2,p_2^2,p_3^2,m_4^2;s_{12},s_{13};0,m_2^2,0,m_4^2)$
\item[-] Box 16: $I_4^D(m_2^2,p_2^2,p_3^2,m_4^2;s_{12},s_{13};0,m_2^2,m_3^2,m_4^2)$
\end{enumerate}
In addition, there are finite integrals, which are included in the implementation.
In the following, we will present a validation of the Kokkos implementation
of these integrals, as well as the simpler tadpole and  bubble integrals.
We also assess code performance, comparing the CPU and GPU implementations
across computing architectures.

\section{Implementation and cross-checks}
\label{sec:impl}

The Kokkos-based implementation is designed as a portable GPU-oriented library in which one-loop scalar integrals are exposed directly as device-executable kernels. In contrast to the object-oriented CPU framework of QCDLoop 2.0, this design does not rely on cache layers or automatic topology triggering through class dispatch. Instead, the interface is intentionally explicit: each topology is invoked through its corresponding kernel entry point. This choice keeps the execution model simple and backend-agnostic, and maps naturally to large-scale parallel evaluation over independent phase-space points.

\subsection{Portable kernel design}

All implementation components are organized in the \texttt{ql::} namespace. The integral kernels and the auxiliary math/utility routines used along the compute path are declared as \texttt{KOKKOS\_INLINE\_FUNCTION}, enabling host/device portability under a single-source programming model. Numerical constants required by the kernels are exposed through device-callable \texttt{Constants<T>} accessors, including scalar constants and coefficient tables used by special-function approximations. The default production path in this work is double precision; support for quadruple precision is available.

\begin{figure*}
    \centering
    \includegraphics[width=\textwidth]{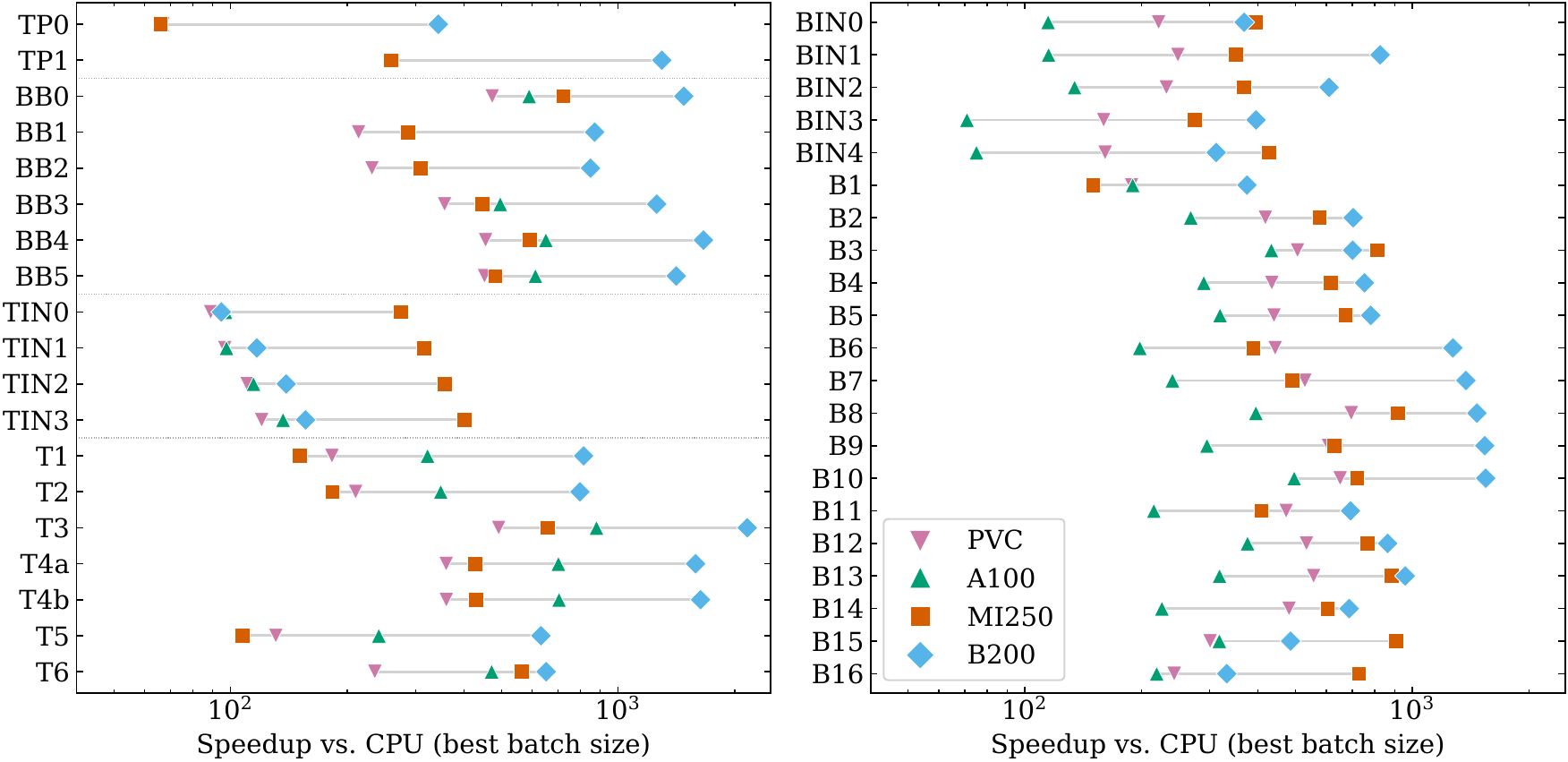}
    \caption{Speedup of GPU architectures over a serial CPU
    baseline for all QCD loop integrals evaluated with Kokkos,
    grouped by integral topology (tadpoles, bubbles,
    triangles, boxes in the notation of the original
    CPU implementation~\cite{Carrazza:2016gav}).
    Each row corresponds to one integral; markers
    show the best achievable speedup across all tested batch sizes.}
    \label{fig:performance}
\end{figure*}

\subsection{Validation}
Numerical correctness is established by direct comparison against the CPU implementation of QCDLoop~\cite{Carrazza:2016gav} over representative kinematic configurations spanning all supported one-loop topologies. For each test point, we compare the Laurent coefficients returned by the GPU kernels to the corresponding CPU reference values, preserving the native QCDLoop coefficient ordering.

The validation includes finite and divergent configurations for tadpole, bubble, triangle, and box families. For each integral, \(10^5\) random input points are generated using ranges of masses and scalar invariants that would appear in typical calculations for LHC phenomenology. Using exactly the same inputs, baseline values are produced with the original CPU implementation in quadruple precision.

For the Kokkos-enabled implementation in double precision, the 64-bit outputs are stored in hexadecimal form to avoid precision loss due to decimal \texttt{printf}-style formatting. Accuracy is then quantified coefficient-by-coefficient in the $\eps$ expansion as the deviation from the quadruple-precision baseline and converted into the number of precise decimal digits, capped at 33 digits for quadruple precision and 16 digits for double precision.

As a first implementation-level validation, results obtained with the Kokkos Serial backend are used as the primary consistency check. Only very small differences are observed in isolated cases where \texttt{Kokkos::log(Kokkos::complex<double>)} is evaluated through the native Kokkos implementation instead of forwarding to \texttt{std::log(std::complex<double>)}.

Finally, cross-architecture accuracy is evaluated on NVIDIA A100 and B200, AMD MI250, and Intel PVC GPUs, allowing a direct assessment of numerical consistency across vendor backends.
Furthermore, we test numerical agreement and accuracy by comparing to Collier~\cite{Denner:2016kdg} and find agreement, as well as a similar numerical stability.

\begin{figure*}
    \centering
    \includegraphics[width=0.9\linewidth]{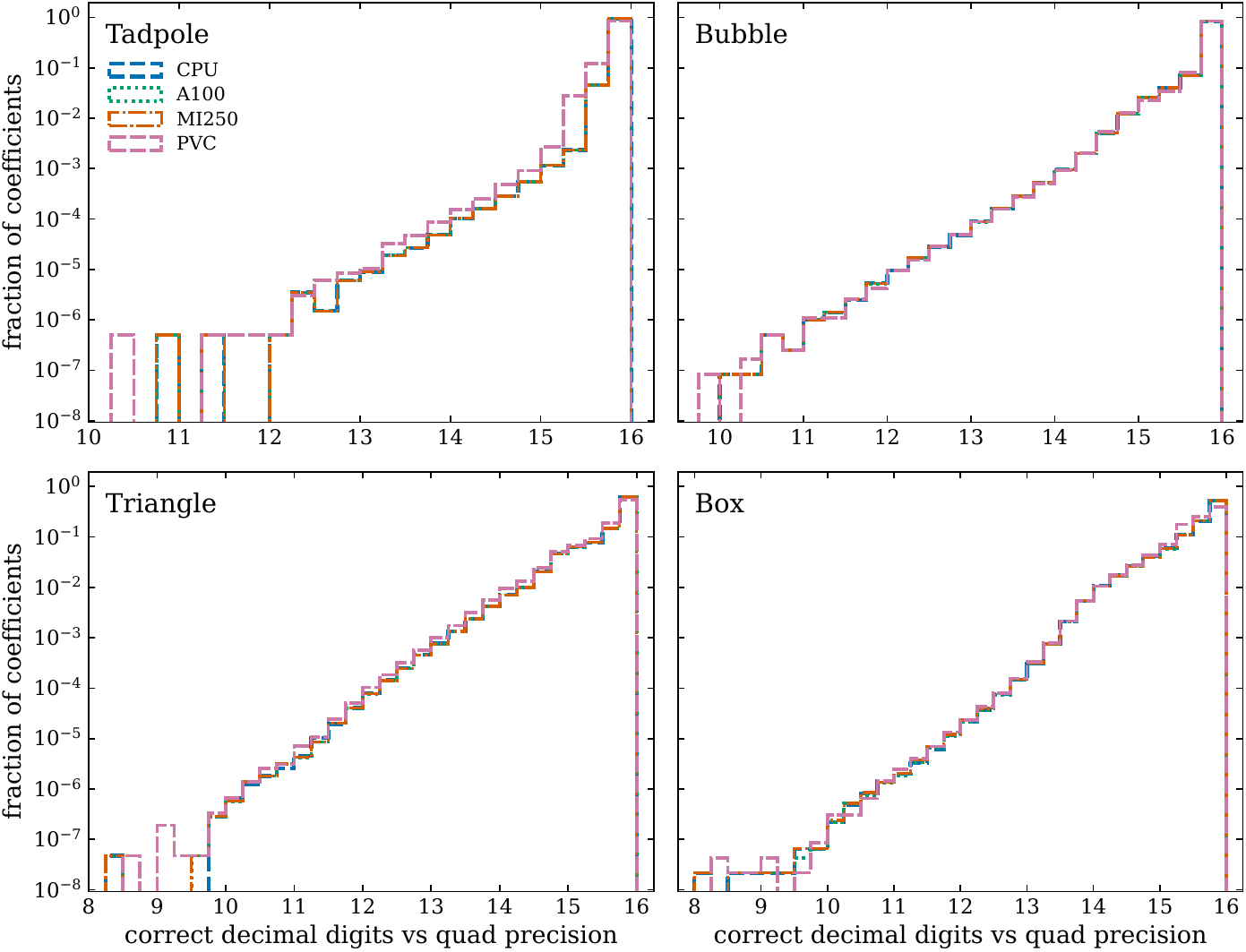}
  \caption{Numerical accuracy of the Kokkos-enabled QCDLoop library in double 
  precision. For each Laurent coefficient $c$ in the $\eps$ expansion we show the
  number of correct decimal digits,
  $d=-\log_{10}\left(|c-c_{\rm ref}|/|c_{\rm ref}|\right)$, relative to the
  quad-precision reference $c_{\rm ref}$ obtained at the same phase-space point.
  The value is capped at $16$, the double-precision limit, so that coefficients
  reproducing the reference to the last available bit populate the rightmost bin.
  \label{fig:accuracy}}
\end{figure*}

Figure~\ref{fig:accuracy} shows the numerical accuracy of the Kokkos-enabled
library in double precision, scored against the quad-precision reference. For
every topology the distribution is peaked at the double-precision limit of 16
accurate digits. The left tail is populated by the numerically unstable regions
of phase space, where cancellations between individually large contributions
reduce the achievable precision. This property of the integrals is therefore
common to all architectures and we observe no significant difference between CPU
and GPU implementations.

\section{Parallel performance across GPU architectures}
\label{sec:perf}

The crucial benchmark of the Kokkos-enabled QCDLoop implementation evaluates
its parallel performance across multiple GPU architectures. We compare batched
evaluations of tadpole, bubble, triangle, and box integrals under a common
setup, and we discuss how performance varies with topology complexity and
hardware characteristics.

In all benchmarks, the reported kernel time measures only the \texttt{Kokkos::parallel\_for} execution (including synchronization) and excludes host-device data transfers. This choice reflects the intended usage in a production-level environment, where integral kernels operate on data already resident on device memory and produce outputs consumed directly by downstream device-resident kernels. 

\noindent \textbf{Experimental Setup:} We evaluate the Kokkos port of QCDLoop across four hardware backends: a
multicore CPU (serial execution), NVIDIA GPUs (A100 and B200), an AMD GPU
(MI250), and an Intel GPU (Ponte Vecchio, PVC). This spans the three major
vendor programming models---CUDA, HIP, and SYCL---exercised through a single
Kokkos~\cite{kokkos} source base, allowing us to assess performance portability
without modifying the physics kernels.

\paragraph{Software environment.}
All configurations use a common toolchain: GCC~13.3.0 as the host compiler,
CMake~3.28.3, and Kokkos~4.7.01 built in \texttt{Release} mode against the
C++17 standard. Each backend is enabled through the corresponding Kokkos
execution space (\texttt{SERIAL}, \texttt{CUDA}, \texttt{HIP}, or \texttt{SYCL})
and target-architecture flag, and is compiled with the vendor compiler and
runtime module summarized in Table~\ref{tab:build-config}. The NVIDIA builds
use \texttt{nvcc} as the device compiler (with \texttt{g++} as host) and enable
CUDA lambdas, \texttt{constexpr} support, and non-fast-math semantics; the AMD
build uses \texttt{hipcc}; and the Intel build uses the oneAPI \texttt{icx}/\texttt{icpx}
compilers provided by the \texttt{frameworks} module on ALCF's Aurora supercomputer.

To ensure numerical reproducibility across backends, the Intel PVC build
additionally compiles both Kokkos and QCDLoop with
\texttt{-fp-model=precise} and \texttt{-ffp-contract=off}, which disable
value-unsafe floating-point reassociation and fused multiply-add (FMA)
contraction so that device results agree with a non-FMA reference
implementation.

\paragraph{Hardware.}
All experiments were run on a single node. The serial CPU baseline is measured
on a single core of an AMD Zen2 host processor; note that the build passes no
Kokkos architecture flag (\texttt{NONE} rather than \texttt{Kokkos\_ARCH\_ZEN2}),
so no Zen2-specific code generation is requested. GPU results are collected on
the NVIDIA A100, NVIDIA B200, AMD MI250, and Intel PVC, with the Intel PVC
results obtained on the ALCF Aurora supercomputer.

\begin{table*}
  \centering
  \setlength{\tabcolsep}{5pt}
  \begin{tabular}{llllll}
    \toprule
    \textbf{Target} & \textbf{Host/Device} & \textbf{Compiler} & \textbf{Vendor} & \textbf{Kokkos} & \textbf{Kokkos} \\
     & \textbf{compiler} & \textbf{version} & \textbf{module} & \textbf{backend} & \textbf{arch.\ flag} \\
    \midrule
    Serial CPU  & \texttt{gcc} / \texttt{g++}          & GCC 13.3.0                  & ---                         & \texttt{Kokkos\_ENABLE\_SERIAL} & \texttt{NONE}\,$^{d}$ \\
    NVIDIA A100 & \texttt{gcc} / \texttt{g++}\,$^{a}$  & CUDA 12.9.1                 & \texttt{cuda/12.9.1}        & \texttt{Kokkos\_ENABLE\_CUDA}   & \texttt{Kokkos\_ARCH\_AMPERE80} \\
    NVIDIA B200 & \texttt{gcc} / \texttt{g++}\,$^{a}$  & CUDA 12.9.1                 & \texttt{cuda/12.9.1}        & \texttt{Kokkos\_ENABLE\_CUDA}   & \texttt{Kokkos\_ARCH\_BLACKWELL100} \\
    AMD MI250   & \texttt{hipcc} / \texttt{hipcc}      & ROCm 7.0.2                  & \texttt{rocm/7.0.2}         & \texttt{Kokkos\_ENABLE\_HIP}    & \texttt{Kokkos\_ARCH\_VEGA90A} \\
    Intel PVC   & \texttt{icx} / \texttt{icpx}         & oneAPI 2025.3.2\,$^{b}$     & \texttt{frameworks}\,$^{c}$ & \texttt{Kokkos\_ENABLE\_SYCL}   & \texttt{Kokkos\_ARCH\_INTEL\_PVC} \\
    \bottomrule
  \end{tabular}
  \caption{Compilers, runtime modules, and Kokkos architecture flags used to build
    QCDLoop for each backend. All builds use the GCC~13.3.0 host toolchain,
    CMake~3.28.3, and Kokkos~4.7.01 (Release, C++17).
    \newline\newline
    \footnotesize
    $^{a}$~The CUDA backend (A100 and B200) uses \texttt{nvcc} as the device
    compiler with \texttt{g++} as the host compiler, with
    \texttt{Kokkos\_ENABLE\_CUDA\_LAMBDA} and \texttt{Kokkos\_ENABLE\_CUDA\_CONSTEXPR}
    enabled and \texttt{Kokkos\_ENABLE\_CUDA\_FASTMATH} disabled.
    $^{b}$~Intel oneAPI DPC++/C++ Compiler 2025.3.2 (build 2025.3.2.20260112). The
    PVC build additionally uses \texttt{-fp-model=precise -ffp-contract=off}.
    $^{c}$~On the ALCF Aurora system (software stack 26.26.0) the compilers are
    provided by the \texttt{frameworks} module (\texttt{aurora\_frameworks-2025.3.1}).
    $^{d}$~The serial CPU host is an AMD Zen2 processor, but the build passes no
    architecture flag (\texttt{NONE}); \texttt{Kokkos\_ARCH\_ZEN2} was not used.}
    \label{tab:build-config}
\end{table*}

Figure~\ref{fig:performance} shows speedups compared to CPU evaluation
for a number of different GPU architectures. We observe that in most cases
the MI250 provides the best performance, and that the acceleration is greater
than a factor 100 for the integrals which are the most time consuming.
We note, that since we did not port the caching algorithm employed in the
original QCDLoop, the speedup seen in a realistic Monte Carlo calculation, where
repeated invariants can be cached, will be smaller than the one presented here.

\section{Outlook}
\label{sec:outlook}
We have presented the first performance-portable implementation of scalar
one-loop Feynman integrals for numerical computations in collider physics.
The computing solutions that we provide are based on
the well established QCDLoop package and are suitable
for both CPU and GPU architectures. It is expected that continued support of
the Kokkos portability framework will enable the new code to be used throughout
the high-luminosity LHC era, during the EIC era, and during the design stage
of a potential future collider. For the future we expect this to be further enabled via
approximate higher-precision availability within Kokkos.
The implementation will be made publicly available on GitHub\footnote{\url{https://github.com/ReetBarik/qcdloop-kokkos.git}}.

\section*{Acknowledgments}
\noindent
We thank Giulia Zanderighi and Keith Ellis for discussions and comments on the draft. 
This manuscript has been authored by Fermi Forward Discovery Group, LLC
under Contract No. 89243024CSC000002 with the U.S.\ Department of Energy,
Office of Science, Office of High Energy Physics.
This work is supported by US Department of Energy, Office of Science, Office of High Energy Physics under the High Energy Physics Center for Computational Excellence (HEP-CCE) under B\&R KA2401045, a collaboration between Argonne National Laboratory, Brookhaven National Laboratory, Fermi National Accelerator Laboratory, Lawrence Berkeley National Laboratory, Oak Ridge National Laboratory, and SLAC National Accelerator Laboratory.
The work of M.K.\ was supported by the U.S. Department of Energy,
Office of Science, Office of Advanced Scientific Computing Research,
Scientific Discovery through Advanced Computing (SciDAC-5) program,
grant “NeuCol”.

This research used resources of the Argonne Leadership Computing Facility,
which is a U.S. Department of Energy Office of Science User Facility
operated under contract DE-AC02-06CH11357. The authors appreciate
the computing resources provided and operated by the Joint Laboratory
for System Evaluation (JLSE) at Argonne National Laboratory.

The authors acknowledge the use of Claude Opus 5.0 to help with debugging and code review. All AI-assisted text was reviewed, edited, and approved by the authors, who take full responsibility for the content.

\appendix
\begin{widetext}
\section{Structure of the interface}
\label{sec:app}

\setcounter{subsubsection}{0}

\vspace{0.5em}
\refstepcounter{subsubsection}
\noindent\textbf{\thesubsubsection\quad Namespace, device functions, and constants}\par
All components are implemented in the \texttt{ql::} namespace. The integral kernels and auxiliary math/utility routines used in the compute path are declared as \texttt{KOKKOS\_INLINE\_FUNCTION}, enabling direct use in host and device execution contexts under a single-source model.

Numerical constants required by the kernels are provided through device-callable \texttt{Constants<T>} accessors in \texttt{ql::}. This includes scalar constants (e.g. \(\pi\), numerical cutoffs, and small-\(\epsilon\) thresholds) as well as coefficient tables used by special-function approximations (Chebyshev and Bernoulli coefficients). This layout ensures that constants are available consistently inside device code without host-only dependencies.

\vspace{0.5em}
\refstepcounter{subsubsection}
\noindent\textbf{\thesubsubsection\quad Precision model}\par
The default production path is double precision. What follows is a brief description of the support for quadruple precision available in the implementation: 
To provide a quad-precision (FP128, 128-bit) evaluation path for QCDLoop on
GPUs, we build on the NVIDIA CUDA Math API FP128 functions~\cite{cuda-math-fp128},
available in CUDA~13 on devices of compute capability $\geq 10.0$. This API is,
however, a purely \emph{scalar} interface: it exposes only device-side
elementary operations on the native \texttt{\_\_float128}/\texttt{\_Float128}
type through functions of the form \texttt{\_\_nv\_fp128\_add}, \texttt{\_mul},
\texttt{\_div}, \texttt{\_sqrt}, \texttt{\_log}, \texttt{\_atan}, and so on. It
provides neither operator syntax, nor a complex-number type, nor the type
traits that the Kokkos performance-portability layer requires in order to store
and manipulate such values in its data structures. Bridging this gap required a
dedicated engineering layer, which we describe here. \\

\noindent\textbf{Scalar wrapper with device/host operator dispatch:}
The raw FP128 type cannot be used with natural arithmetic syntax on the device,
since each operation must be routed through an explicit \texttt{\_\_nv\_fp128\_*}
intrinsic. We therefore introduce a thin wrapper \texttt{struct fp128\_wrapper}
that stores a single \texttt{\_\_float128} member and overloads the full set of
arithmetic (\texttt{+}, \texttt{-}, \texttt{*}, \texttt{/}), compound-assignment,
unary, and comparison operators. Each operator is compiled twice: inside a
device context (\texttt{\_\_CUDA\_ARCH\_\_}) it dispatches to the corresponding
\texttt{\_\_nv\_fp128\_*} intrinsic, while on the host it falls back to the
native compiler-supported \texttt{\_\_float128} operators. All operators are
annotated \texttt{KOKKOS\_INLINE\_FUNCTION} so that a single source expression
compiles correctly for both the host and device back-ends. Mixed-type operators
(e.g. \texttt{fp128\_t}~$\times$~\texttt{double}) are provided as
SFINAE-constrained templates that promote the operand to full precision,
mirroring the implicit-promotion semantics of the standard library and of
\texttt{Kokkos::complex}. An implicit conversion back to the underlying scalar
keeps the wrapper a drop-in replacement wherever \texttt{\_\_float128} is
expected. \\

\noindent\textbf{Custom quad-precision complex type:}
The choice to instantiate QCDLoop's kernels with a complex-number type of quad 
precision with \texttt{Kokkos::complex<\_\_float128>} is not a viable one because
\texttt{Kokkos::complex<T>} requires its value type \texttt{T} to expose native 
arithmetic and the associated arithmetic traits, which the device-only FP128 scalar 
(whose operations must be funneled through intrinsics) does not satisfy. Rather than
specialize the full set of Kokkos traits for a device type that cannot support
them natively, we implement a minimal \texttt{struct quad\_complex} holding two
\texttt{fp128\_t} components. On top of it we provide exactly the operations the
kernels use: complex \texttt{+}, \texttt{-}, \texttt{*}, \texttt{/},
scalar--complex mixed operators, \texttt{abs}, \texttt{conj},
and branch-stable \texttt{log} and \texttt{sqrt}.
The complex square root in particular reproduces the Kokkos numerical strategy
of selecting between the \texttt{re > 0} and \texttt{re \(\leq\) 0} branches to
avoid catastrophic cancellation, which is essential for preserving the tiny
imaginary parts introduced by the $i\varepsilon$ prescription. \\

\noindent With the above in place, both \texttt{fp128\_t} and \texttt{quad\_complex} can be used
directly as \texttt{Kokkos::View} element types (e.g.
\texttt{Kokkos::View<fp128\_t*>} and \texttt{Kokkos::View<quad\_complex*[3]>}),
so that the existing templated QCDLoop kernels compile unchanged simply by
substituting the quad-precision types for the double-precision ones. Finally, 
we note that this quad-precision mode is hardware-specific:
because it relies on the CUDA FP128 device intrinsics, it is available only on
NVIDIA GPUs of compute capability $\geq 10.0$ (the Blackwell generation and
later), unlike the double-precision path which remains portable across all
Kokkos back-ends.

\vspace{0.5em}
\refstepcounter{subsubsection}
\noindent\textbf{\thesubsubsection\quad Kernel interface and argument structure}\par
The four supported topologies are exposed through topology-specific kernels:
\texttt{TP} (tadpole), \texttt{BB} (bubble), \texttt{TR} (triangle), and \texttt{BO} (box). All four kernels share the same logical argument ordering,
\[
\texttt{\{TP,BB,TR,BO\}(result\_view, mu2\_view, mass\_view, invariant\_view, i)}.
\]
For each phase-space point \(i\), the inputs are the renormalization scale \(\mu^2\), internal masses, and kinematic invariants. The topology-dependent input dimensions are: tadpole \((1,0)\), bubble \((2,1)\), triangle \((3,3)\), and box \((4,6)\), corresponding to (number of masses, number of invariants), cf.\ Eqs.~\eqref{eq:topologies} and~\eqref{eq:topologies_2}.

\vspace{0.5em}
\refstepcounter{subsubsection}
\noindent\textbf{\thesubsubsection\quad Output convention and batched execution}\par
The output is a result view with three Laurent coefficients per phase-space point, preserving the original QCDLoop ordering:
\[
\texttt{res[0]}:\epsilon^{0}\ \text{(finite part)},\qquad
\texttt{res[1]}:\epsilon^{-1},\qquad
\texttt{res[2]}:\epsilon^{-2}.
\]
In practical usage, a batch of \(N\) phase-space points is processed by launching the selected topology kernel over \(i=0,\dots,N-1\). This organization provides a uniform physics interface while exposing parallelism over independent phase-space points for performance-portable execution across backends.

\end{widetext}

\bibliography{main}
\end{document}